\documentclass[review]{elsarticle}
\biboptions{sort&compress}

\usepackage{amsmath}
\usepackage{amssymb}
\usepackage{graphicx}
\usepackage{subcaption}
\usepackage{bm}
\usepackage{cases}

\newcommand{\A}{\bm{A}}
\newcommand{\R}{\bm{R}}
\renewcommand{\a}{\bm{a}}
\renewcommand{\u}{\bm{u}}
\newcommand{\x}{\bm{x}}

\journal{Elsevier journal}

\begin{document}

\begin{frontmatter}



\title{Design and implementation of net zero displacement filter for the synthesis of a mechanical shock signal under specified shock response spectrum}


\author{Yinzhong Yan}
\author{Q.M. Li\corref{cor1}}
\cortext[cor1]{Corresponding author.}
\ead{qingming.li@manchester.ac.uk}

\address{Department of Mechanical, Aerospace and Civil Engineering, School of Engineering, The University of Manchester, Manchester M13 9PL, United Kingdom}

\begin{abstract}

Electronic and optical products are vulnerable under mechanical shock environment.
Designed products need to be validated by shock testing and/or numerical simulation, using representative acceleration-time history signals.
However, specifications derived from measurements are normally given in terms of shock response spectrum (SRS) without a corresponding time history signal, and therefore, there is a need to synthesis acceleration-time histories from a given SRS specification.
This paper proposed a net zero displacement filter and a realistic time history synthesis method.
By scaling a relevant field measurement, acceleration-time histories can be synthesized, which can meet the net zero displacement constrain and a given SRS specification within $\pm$3 dB margin.

\end{abstract}


\end{frontmatter}


\section{Introduction}

The shock response spectrum (SRS) has been widely used as the testing specification tool by various standards to describe the severities of different kinds of shocks\cite{ECSS2015, 810g, 901e}.
In many cases, a designed product needs to be validated under a derived specification without a corresponding time history of the shock signal.
However, there is no bijective relationship between shock acceleration-time history and its SRS\cite{ECSS2015}, which implies that a specific SRS curve may correspond to different shock acceleration-time histories. 
A method to synthesise more `realistic' acceleration-time histories that meet given SRS specification is necessary, which is particularly important for the use of electrodynamic shaker testing and numerical simulation to generate stimulant shocks\cite{lalanne2013mechanical}.

The method adopted commonly in shock synthesis is the linear combination of limited waveform bases, e.g., damped sine wave\cite{smallwood1975time}, Kern and Hayes' function\cite{kern1984transient}, ZERD function\cite{fisher1977digital}, WAVSIN waveform\cite{yang1972development}, and wavelet\cite{brake2011inverse}. 
Recently, many SRS synthesis methods were proposed with the help of optimization algorithms.
For example, Brake\cite{brake2011inverse} used several basic waveforms and genetic algorithm (GA); Hwang and Duran\cite{hwang2016stochastic} synthesised shock signal with damped sine waves and Monte Carlo simulation; Monti and Gasbarri\cite{monti2017dynamic} used damped sine wave and GA.
Although these methods can synthesis acceleration-time histories to satisfy a given SRS specification, they ignored the intrinsic `net zero displacement' requirement for a shock signal\cite{ECSS2015, lalanne2013mechanical}, which may lead to some practical difficulties in implementing the synthesized shock signal into a shock generator.
A time-delay, which is inversely proportional to the frequency of the wavelet, was introduced in a wavelet-based shock synthesis algorithm in \cite{irvineShock}.
This algorithm has been adopted by both ESA's and NASA's documentations\cite{ECSS2015,Ferebee2008a}.
The synthesized acceleration signals based on the time-delayed wavelet algorithm can meet the `net zero displacement' requirement but may have significantly different temporal structure and severity from those of the real field shocks\cite{hwang2016stochastic}.

As an impulsive response, a shock is largely determined by the transmission structure.
According to the `similarity-heritage-extrapolation' method\cite[p.~75]{ECSS2015}, a possibly different, but physically similar, shock environment from an unknown structure can be evaluated by using existing field measurements from a similar structure.
Following this idea, this study extracts physical information from a structure by applying a net zero displacement filter (NZDF) bank on a representative field shock signal.
Filtered results are scaled and reconstructed with particle swarm optimization (PSO) algorithm to synthesize shock signals.
The reconstructed signal can satisfy a given SRS specification while meeting the net zero displacement condition and relating to the physically measured shock signal.

\section{Net Zero Displacement Filter}

\subsection{Net Zero Displacement Condition}

Shock is the response of a structure under an impulsive loading\cite{yan2019low}.
Normally shocks are defined in elastic response domain, which does not damage and/or permanently deform the main structure, e.g., a spacecraft.
The deformation of the structure shall normally return to the equilibrium position at the end of a shock event, which means that there shall be no net velocity and displacement change.
It is more convenient to consider only the net zero displacement condition, since it is also a sufficient condition for the net zero velocity condition.
For an acceleration measurement $\ddot{u}(t)$ of a shock event, the net zero displacement condition can be described by
\begin{equation}\label{net_zero_condition}
u(t)|_{t \rightarrow \infty} = 0
\end{equation}
where $u(t)$ is the displacement function of the shock by integrating $\ddot{u}(t)$ twice
\begin{equation}
u(t)=\int_0^t \bigg(\int_0^{\bar{t}} \ddot{u}(t) \  dt\bigg) d\bar{t}.
\end{equation}

\subsection{Design of Net Zero Displacement Filter}\label{section_design_filter}

Gammatone-like filters\cite{katsiamis2007practical,lyon2017human} have extensive applications in an auditory system, whose impulse response function resembles the Gammatone function
\begin{equation}
g(t)=at^{N-1} e^{-bt} \cos(\omega_c t+\phi),
\end{equation}
where $a$ is the amplitude, $N$ is the order, $b$ is the decay rate, $\omega_c$ is the angular frequency of the carrier wave, and $\phi$ is the phase.

It is worth to note that the Gammatone function is essentially the same as the shock waveform derived based on the response characteristics of a linear elastic structure under impulsive loading in the study of mechanical shock\cite{yan2019general}.
Some of the features of Gammatone function was initially realized in a basilar membrane model by Flanagan\cite{flanagan1960models} in 1960.
The complete features and the definition of Gammatone function were described between 1972 and 1980 by Johannesma\cite{johannesma1972pre} and Aertsen and Johannesma\cite{aertsen1980spectro}.
Since then, Gammatone function has become the basis of many successful studies in audio system modelling\cite{lyon2017human}.
Gammatone function and shock waveform function were realised independently in two different and separated research fields.
The latter was derived analytically with clear physical meanings.

Carrying the characteristics of shock signals, this type of filters is introduced for the design of NZDF to ensure realistically-filtered results for a shock signal.
Filtered acceleration signal $\ddot{u}(t, \omega)$ at centre frequency $\omega$ should satisfy the net zero displacement condition in Eq.(\ref{net_zero_condition}).

The Laplace transform $G(s)$ of Gammatone function $g(t)$ is the transfer function of the Gammatone filter, i.e.,
\begin{equation}
	G(s)=\frac{ e^{j \phi} (s+b+j\omega_c)^N +\\
		e^{-j \phi} (s+b-j \omega_c)^N}{\big( (s+b)^2 + \omega_c^2 \big) ^N}
\end{equation}
where $s$ is a complex number and $j$ is the imaginary unit.
A useful parameter alternation for simplification is to replace $\omega_c$ and $b$ with centre angular frequency $\omega$ and quality factor $Q$, respectively\cite{lyon1997all},
\begin{equation}\label{transfer_function_Q}
\resizebox{!}{!}{$
G(s)=\frac{ e^{j \phi} \big(s+\omega/(2Q)+j\omega \sqrt{1-1/(4Q^2)}\big)^N + \\
	 e^{-j \phi} \big(s+\omega/(2Q)-j\omega \sqrt{1-1/(4Q^2)}\big)^N}{(s^2 +(\omega/Q)s + \omega^2)^N}
$}
\end{equation}
where $\omega=\sqrt{\omega_c^2 + b^2}$ and $Q=\omega/(2b)$.

\begin{figure}
	\centering
	\includegraphics[width=0.55\linewidth]{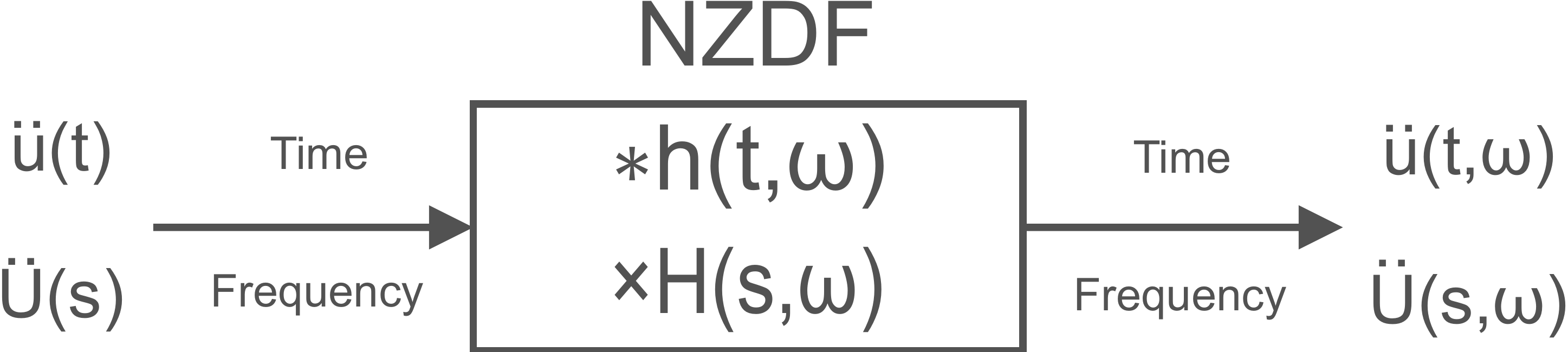}
	\caption{Diagram of NZDF system with centre frequency $\omega$}
	\label{diagram_input_output}
\end{figure}

The all-pole Gammatone filter\cite{slaney1993efficient} (APGF) is defined by discarding the zeros from a pole-zero decomposition of Eq.(\ref{transfer_function_Q}), i.e.,
\begin{equation}
G(s)=\frac{ K }{(s^2 +(\omega/Q)s + \omega^2)^N}
\end{equation}
where $K$ is a constant gain term to be determined in section \ref{section_parameters}.
Based on the expression of APGF, the transfer function $H(s,\omega)$ for NZDF is proposed here to have the following form
\begin{equation}\label{tf_NZDF}
H(s, \omega)=\frac{ K s^M }{(s^2 +(\omega/Q)s + \omega^2)^N},
\end{equation}
where the term $s^M$ ($M$ is a constant to be determined later) is introduced to fine tune the displacement of filtered signal.
In Eq.(\ref{tf_NZDF}), the NZDF transfer function is expressed as $H(s,\omega)$ because it will be used later in a filter bank with various centre frequencies.
The filtered results of $\ddot{u}(t)$ and $u(t)$ can be calculated by the inverse Laplace transform of
\begin{equation}
	\begin{split}
		\ddot{U}(s,\omega)&=\ddot{U}(s)\cdot H(s,\omega)\\
		&=\frac{ K s^M \ddot{U}(s)}{(s^2 +(\omega/Q)s + \omega^2)^N},
	\end{split}
\end{equation}
\begin{equation}\label{displacement_laplace}
	\begin{split}
		U(s,\omega)&=\frac{1}{s^2} \ddot{U}(s,\omega)\\
		&=\frac{ K s^{M-2} \ddot{U}(s)}{(s^2 +(\omega/Q)s + \omega^2)^N},
	\end{split}
\end{equation}
where $U(s,\omega)$, $\ddot{U}(s,\omega)$ and $\ddot{U}(s)$ are the Laplace transforms of $u(t,\omega)$, $\ddot{u}(t,\omega)$ and $\ddot{u}(t)$, respectively.
A diagram of the NZDF system is shown in Fig.\ref{diagram_input_output}.

The range of parameter $M$ can be bounded by applying both initial and final theorems in Laplace transform as shown in Eqs.(\ref{initial_initial_theorem}) and (\ref{final_value}) respectively: the initial value of the impulse response of NZDF, i.e. $h(t,\omega)$, needs to be a finite value; the final value of $u(t,\omega)$ needs to be zero as required in Eq.(\ref{net_zero_condition}),
\begin{numcases}{}
|h(0^+,\omega)|=|\lim\limits_{s \rightarrow \infty} sH(s,\omega)| < C \label{initial_initial_theorem}\\
u(\infty,\omega)=\lim\limits_{s \rightarrow 0} sU(s,\omega)=0 \label{final_value}
\end{numcases}
where $C$ is a positive finite constant, which lead to
\begin{equation}
2 \leq M \leq 2N-1 \label{m_bound}.
\end{equation}
The proof of Eq.(\ref{m_bound}) is given in \ref{proof_appendix}.

A NZDF is defined by its transfer function in the form of Eq.(\ref{tf_NZDF}) with satisfying the inequalities in Eq.(\ref{m_bound}).

\subsection{Choice of parameters for shock events}\label{section_parameters}

NZDF bank contains a class of filters at discrete frequencies in the concerned frequency range.
Each NZDF can be determined by a combinations of parameters ($K, M, N, Q$).
This subsection introduces a set of parameters for shock event to achieve small group delay ($\tau$), narrow bandwidth ($\beta$) and better similarity between synthesized and original shock signals.

The group delay of Gammatone-like filter has been studied previously in Refs.\cite{yan2019general,katsiamis2007practical}, which can be estimated by
\begin{equation}\label{group_delay}
\tau \omega = 2NQ
\end{equation}
Second-order ($N$=2) NZDF is adopted here to minimise filter's group delay and meet Eq.(\ref{m_bound}).
Another reason for the choice of $N$=2 is to have high similarity between its impulse response (close to Kern and Hayes' function\cite{kern1984transient}) and field shock measurements, which can avoid large distortion during filtering process in the temporal domain.

The parameter $M$ mainly influences the phase information but has a limited effect on NZDF in both temporal and frequency domains.
This parameter can be set arbitrary as long as it meets Eq.(\ref{m_bound}).
In the case $N=2$, parameter $M$ could be either 2 or 3.
For the consideration of numerical stability, $M=2$ is adopted to minimize $|h(0^+,\omega)|$.

The constant gain term $K$ is chosen to make the peak gain at centre frequency to be unity, as shown in Eq.(\ref{unity_gain}).
\begin{equation}\label{unity_gain}
|H(j\omega,\omega)|=1 \quad \Rightarrow \quad K= (\frac{\omega}{Q})^2
\end{equation}

\begin{figure}
	\centering
	\includegraphics[width=0.7\linewidth]{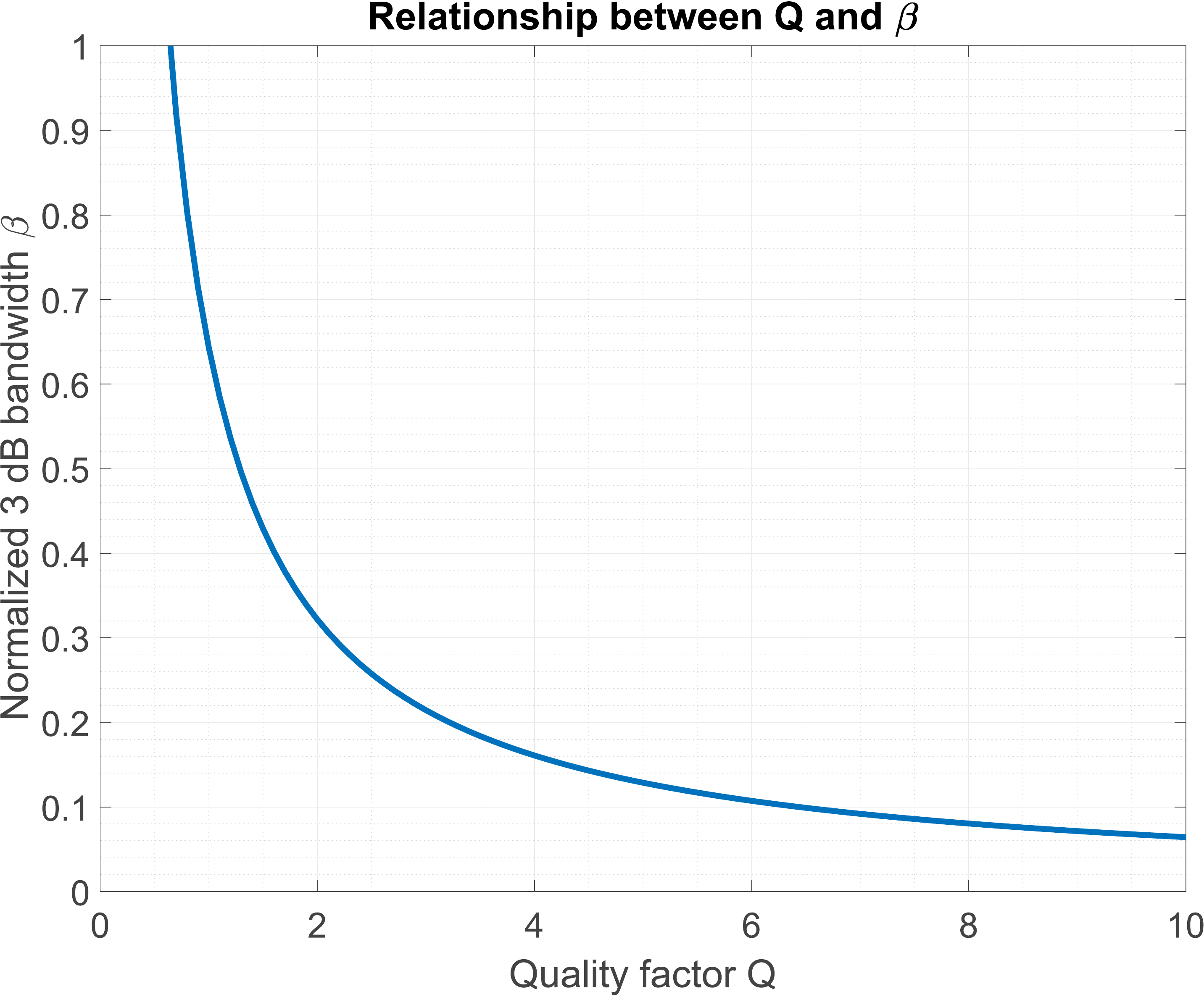}
	\caption{Relationship between the quality factor $Q$ and the normalized bandwidth $\beta$}
	\label{Q_beta_plot}
\end{figure}

The quality factor $Q$ is related to the bandwidth of the filter.
In this study, the 3 dB bandwidth normalized to the centre frequency is introduced and defined as $\beta$.
\begin{equation}\label{normalized_bandwidth}
\beta = \frac{\omega_{\text{UB}} - \omega_{\text{LB}}}{\omega}
\end{equation}
Here the $\omega_{\text{UB}}$ and $\omega_{\text{LB}}$ are the pair of the upper and lower bounds of frequencies where the threshold value is 3 dB lower than the unity (maximum gain of NZDF).
By solving Eq.(\ref{bandwidth_determination1}), the pair of frequencies can be determined.
\begin{equation}\label{bandwidth_determination1}
|H(j\omega_{\text{B}},\omega)| \approx \frac{ 1 }{ \sqrt{2} }
\end{equation}
where B=UB and B=LB are applied.
Using Eq.(\ref{tf_NZDF}), Eq.(\ref{bandwidth_determination1}) leads to
\begin{equation}\label{bandwidth_determination2}
\frac{\omega_{\text{B}}^2 \omega^2}{Q^2 \left(\omega_{\text{B}}^2-\omega^2\right)^2+\omega_{\text{B}}^2 \omega^2}\approx\frac{1}{\sqrt{2}}.
\end{equation}
The solution of Eq.(\ref{bandwidth_determination2}) are
\begin{equation}
\begin{cases}
\omega_{\text{UB}}= \omega \sqrt{\frac{2 Q^2+\sqrt{4 \left(\sqrt{2}-1\right) Q^2-2 \sqrt{2}+3}+\sqrt{2}-1}{2 Q^2}}\\
\omega_{\text{LB}}= \omega \sqrt{\frac{2 Q^2-\sqrt{4 \left(\sqrt{2}-1\right) Q^2-2 \sqrt{2}+3}+\sqrt{2}-1}{2 Q^2}}
\end{cases}.
\end{equation}
By substituting $\omega_{\text{UB}}$ and $\omega_{\text{LB}}$ into Eq.(\ref{normalized_bandwidth}), the relationship between $Q$ and the normalized 3 dB bandwidth $\beta$ is obtained by
\begin{equation}\label{Q_beta_relationship}
\resizebox{0.9\linewidth}{!}{$
\beta=\sqrt{\frac{2 Q^2+\sqrt{4 \left(\sqrt{2}-1\right) Q^2-2 \sqrt{2}+3}+\sqrt{2}-1}{2 Q^2}} - \sqrt{\frac{2 Q^2-\sqrt{4 \left(\sqrt{2}-1\right) Q^2-2 \sqrt{2}+3}+\sqrt{2}-1}{2 Q^2}}
.$}
\end{equation}
This relationship is plotted in Fig.\ref{Q_beta_plot}, which can help to obtain a suitable quality factor $Q$.
For shock synthesis purpose, the bandwidth of NZDF shall be consistent with the spacing of the filter bank introduced in the following section.
In this study, 1/6 octave spacing is adopted as shown in section \ref{section_synthesis}, which is equivalent to a normalized bandwidth $\beta=0.1225$.
From Fig.\ref{Q_beta_plot}, $Q=5$ is selected as the closest integer number, and the NZDF is finalized by
\begin{equation}\label{tf_final}
H(s,\omega)=\frac{ s^2 \omega^2 }{(5s^2 +\omega s + 5\omega^2)^2}.
\end{equation}

\subsection{The properties of NZDF in temporal and frequency domain}

\begin{figure}[t]
	\centering
	\begin{subfigure}{0.49\textwidth}
		\includegraphics[width=\linewidth]{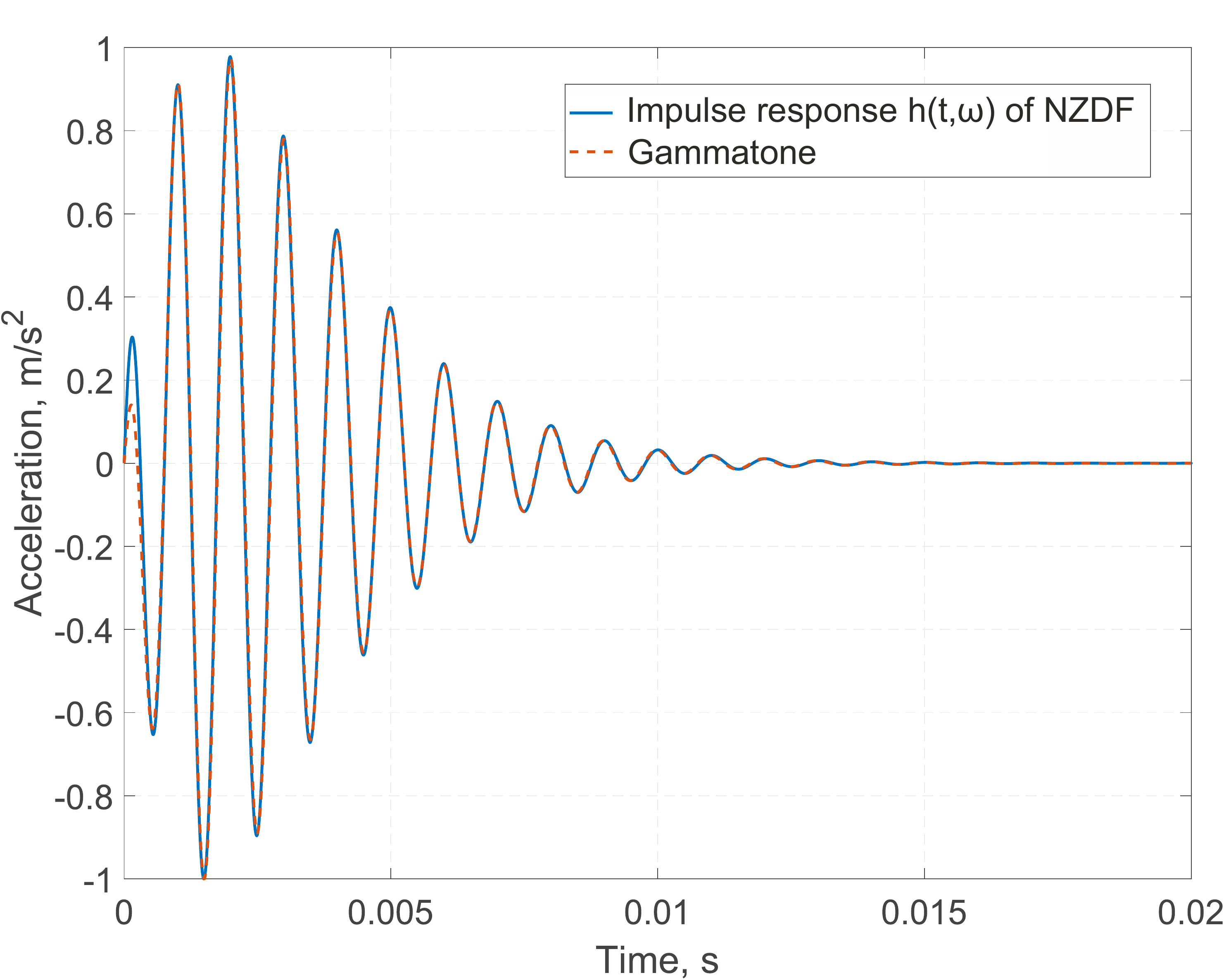}
		\caption{Acceleration}
		\label{Gammtone_h(t)_acc}
	\end{subfigure}
	\begin{subfigure}{0.49\textwidth}
		\includegraphics[width=\linewidth]{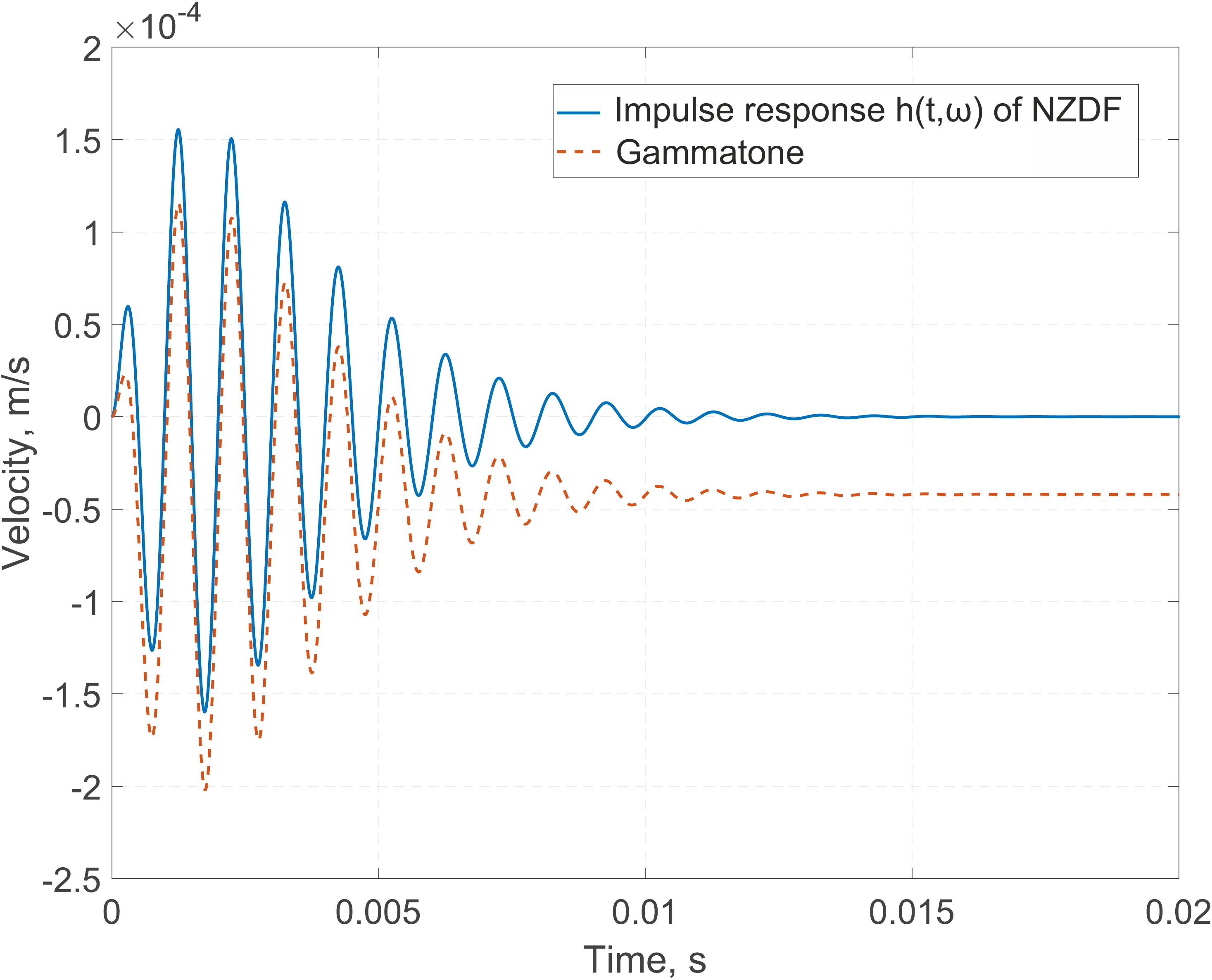}
		\caption{Velocity}
		\label{Gammtone_h(t)_vol}
	\end{subfigure}
	\begin{subfigure}{0.49\textwidth}
		\includegraphics[width=\linewidth]{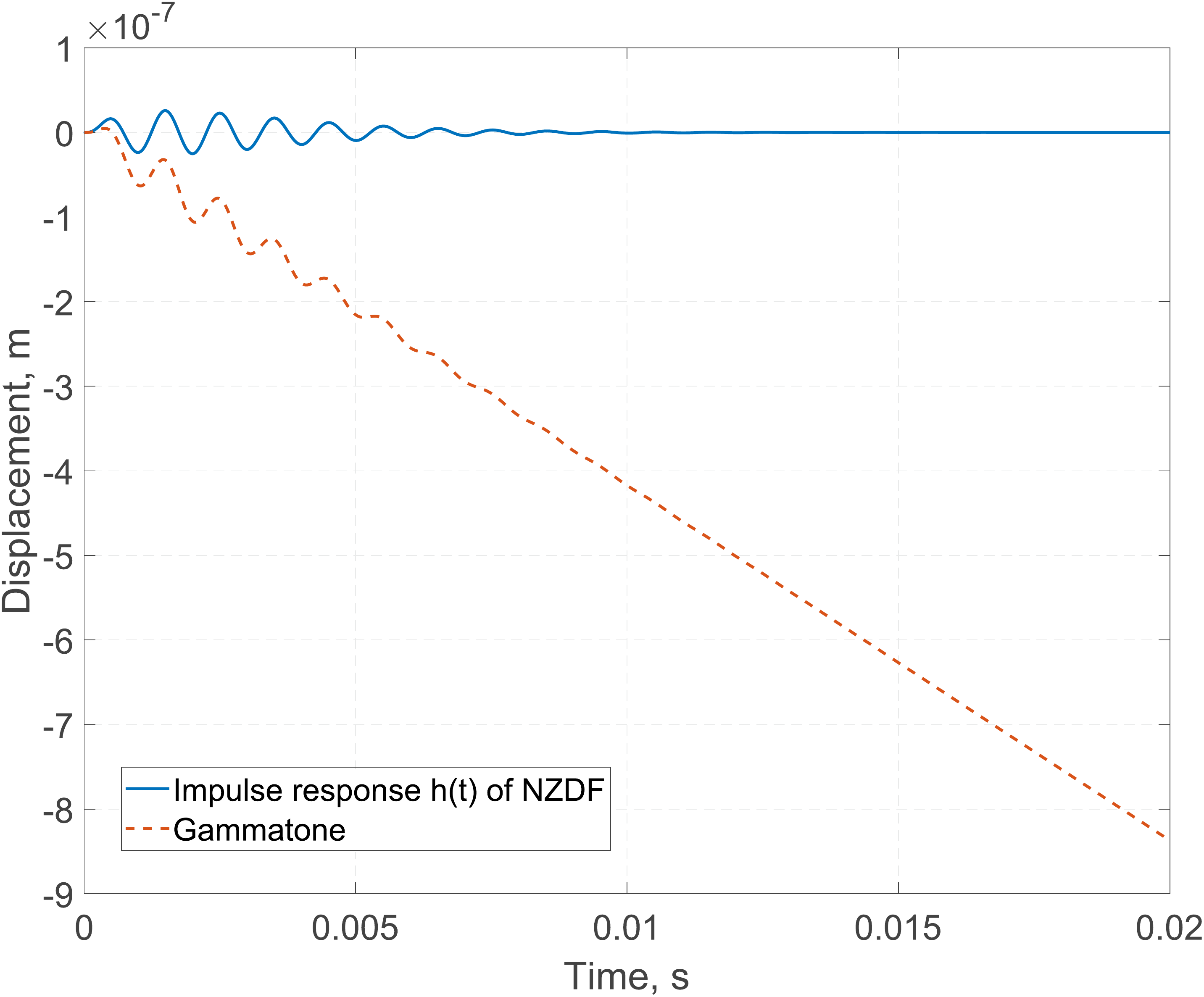}
		\caption{Displacement}
		\label{Gammtone_h(t)_dis}
	\end{subfigure}
	\caption{Acceleration, velocity, and displacement of Gammatone and impulse response $h(t,\omega)$ of NZDF ($\frac{\omega}{2\pi}$=1000 Hz)}
\end{figure}

Regarding the temporal properties of NZDT, its acceleration impulse response $h(t,\omega)$, as well as its first and second order integrations will be damped out and approach to zero.
For comparison, common Gammatone and NZDF's impulse response $h(t,\omega)$ at $\frac{\omega}{2\pi}$=1000 Hz are shown in Fig.\ref{Gammtone_h(t)_acc}, with other parameters given in section \ref{section_parameters}.
Their first and second integrations are also calculated and depicted in Figs.\ref{Gammtone_h(t)_vol} and \ref{Gammtone_h(t)_dis} as velocity and displacement, respectively.
The accelerations of Gammatone and $h(t,\omega)$ are very similar.
The limited difference can only be observed in the first cycle, and both signals finally damped out with the increase of time.
However, after integration to velocity, the Gammatone has an obvious zero shift, which leads to infinite displacement.
While the NZDF's velocity and displacement impulse responses still oscillate around zero and are damped out finally. 

\begin{figure}
	\centering
	\includegraphics[width=0.7\linewidth]{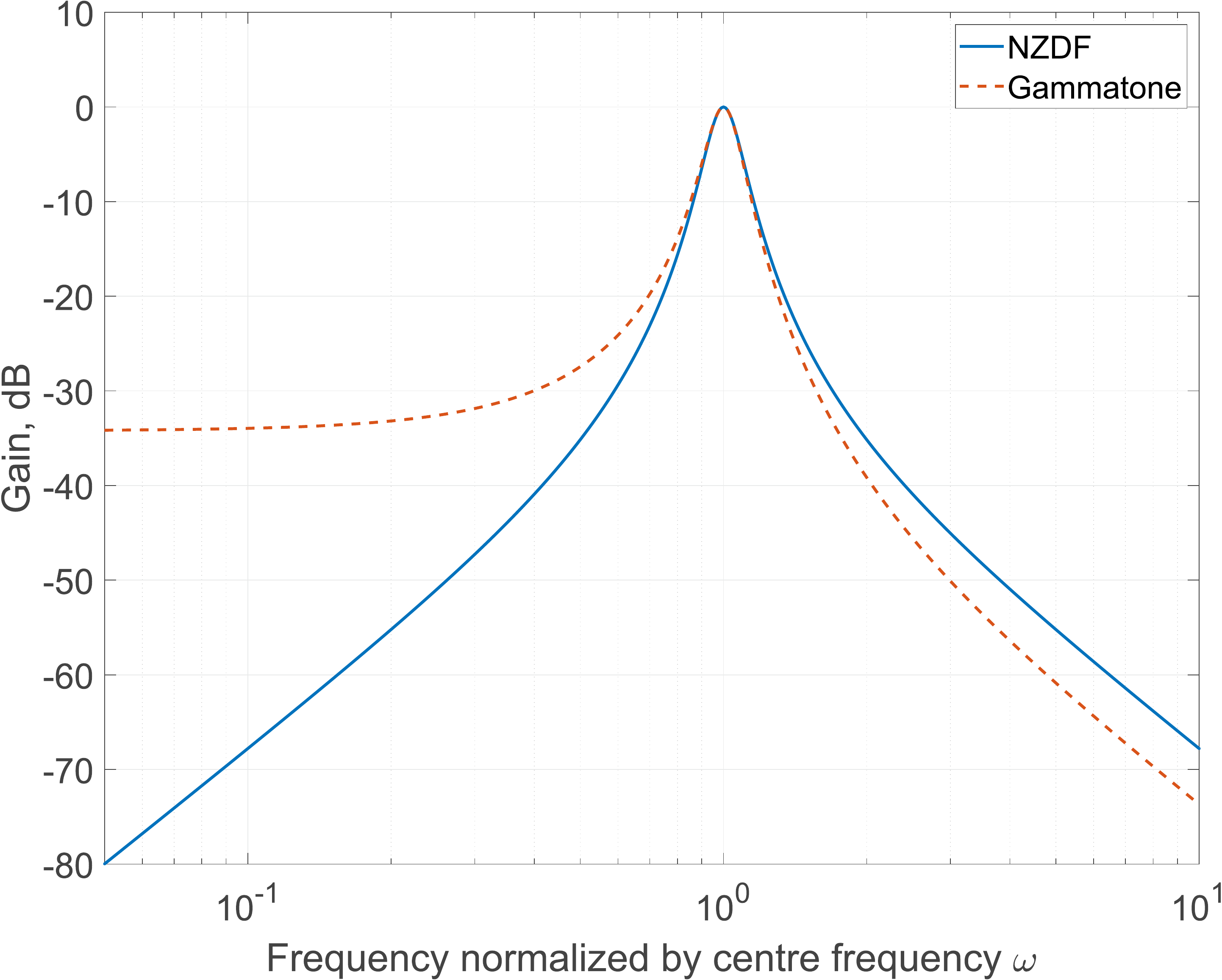}
	\caption{Frequency response of Gammatone and NZDF}
	\label{NZDF_Gamma_frf}
\end{figure}

Fig.\ref{NZDF_Gamma_frf} illustrates the comparison between common Gammatone and NZDF in terms of their frequency responses.
With the same set of parameters, NZDF performs closely to Gammatone around and after centre frequency.
The Gammatone maintains a relatively constant gain level for the low-frequency tail, through which low-frequency oscillation is still likely to be retained.
In contrast, NZDF has a linear (in log scale) low-frequency tail, which can filter out both low- and high-frequency contents but only retain information around the centre frequency.

\section{Shock synthesis}
\label{section_synthesis}

The core idea of this shock synthesis method is to find out a vector space $\mathbb{R}^{n,m}$ with its basis matrix $\A$, so that the SRS of vectors in such vector space can span the entire SRS space. 
In other word, with a given SRS specification, there always exists a linear combination of bases ($\a_i$),
\begin{equation}
\ddot{\u}'=\A \x=\sum_{i=1}^{m} x_i \a_i
\end{equation}
that the SRS of $\ddot{\u}'$ can meet the SRS specification,
\begin{equation}
|SRS(\ddot{\u}')-SRS_{spec}|<\epsilon
\end{equation}
where $\ddot{\u}'$ is the synthesised acceleration of the shock, $\x=[x_1,x_2,x_3,\ldots,x_m]^{\top}$ is the coefficient vector, basis $\a_i$ is the $i$th column vector of $\A$ with $n$ samples, $SRS(\cdot)$ is the SRS algorithm, $SRS_{spec}$ is the SRS testing specification and $\epsilon>0$ is the tolerance (usually 3 dB in shock testing standards).

The basis matrix $\A$ is constructed by passing a field shock measurement $\ddot{\u}$ through a series of NZDF at equal spacing frequencies,
\begin{equation}
\a_i=\ddot{\u}(\omega_i)
\end{equation}
where the vector $\ddot{\u}(\omega_i)$ is the discrete form of $\ddot{u}(t, \omega_i)$ in terms of $t$.
In this way, the synthesised shock $\ddot{\u}'$ also has a net zero displacement change since the cumulative sum (integral) is a linear operator.
It is suggested that a similar field shock $\ddot{\u}$ should be used for the best synthesis performance.
More specificity, field shocks from already tested structures presenting similar architecture, design, configuration and under similar shock generating mechanism are preferable.
In general, shocks measured from laboratory testing (e.g. generated from metal-metal impact) are also acceptable, as their waveforms are more relevant to mechanical shocks compared to other basic waveforms, e.g., damped sines, or wavelets.

The frequency spacing can be decided according to practical need.
To be consistent and comparable with the current state-of-art method\cite{irvineShock}, frequencies at every 1/6 octave is adopted, i.e.,
\begin{equation}
\frac{\omega_{i+1}}{\omega_{i}}=2^{\frac{1}{6}}\approx 1.1225,
\end{equation}
which is roughly equivalent to a normalized bandwidth $\beta$ at 0.1225.

The coefficient vector $\x$ can be obtained by solving the minimization problem in Eq.(\ref{objective_function}) with existing optimization algorithm, e.g., PSO, simulated annealing or genetic algorithm,
\begin{equation}\label{objective_function}
 \underset{\x\in\R^m}{\arg\min} \  \| \log_{10}(SRS(\A \x)) - \log_{10}(SRS_{spec}) \|
\end{equation}
where the $\| \cdot \|$ returns the common Euclidean norm of a vector.
In this study, PSO algorithm from Matlab is used for this purpose.
All the bases come from NZDF with different centre frequencies, which are approximately but not strictly independent and orthogonal.
Thus, this optimization process tends to converge to a unique solution with less time-consuming. 

\section{Case study}

\begin{table}
	\centering
	\caption{Example SRS specification from ECSS shock handbook\cite{ECSS2015}}
	\label{SRS_spec}
	\small
	\begin{tabular}{c c}
		\hline 
		Natural Frequency (Hz) & Peak Acceleration (m/s$^2$)\\ 
		\hline 
		100 & 300\\
		1800 & 10000\\
		10000 & 10000\\
		\hline
	\end{tabular}
\end{table}

\begin{figure}
	\centering
	\includegraphics[width=0.7\linewidth]{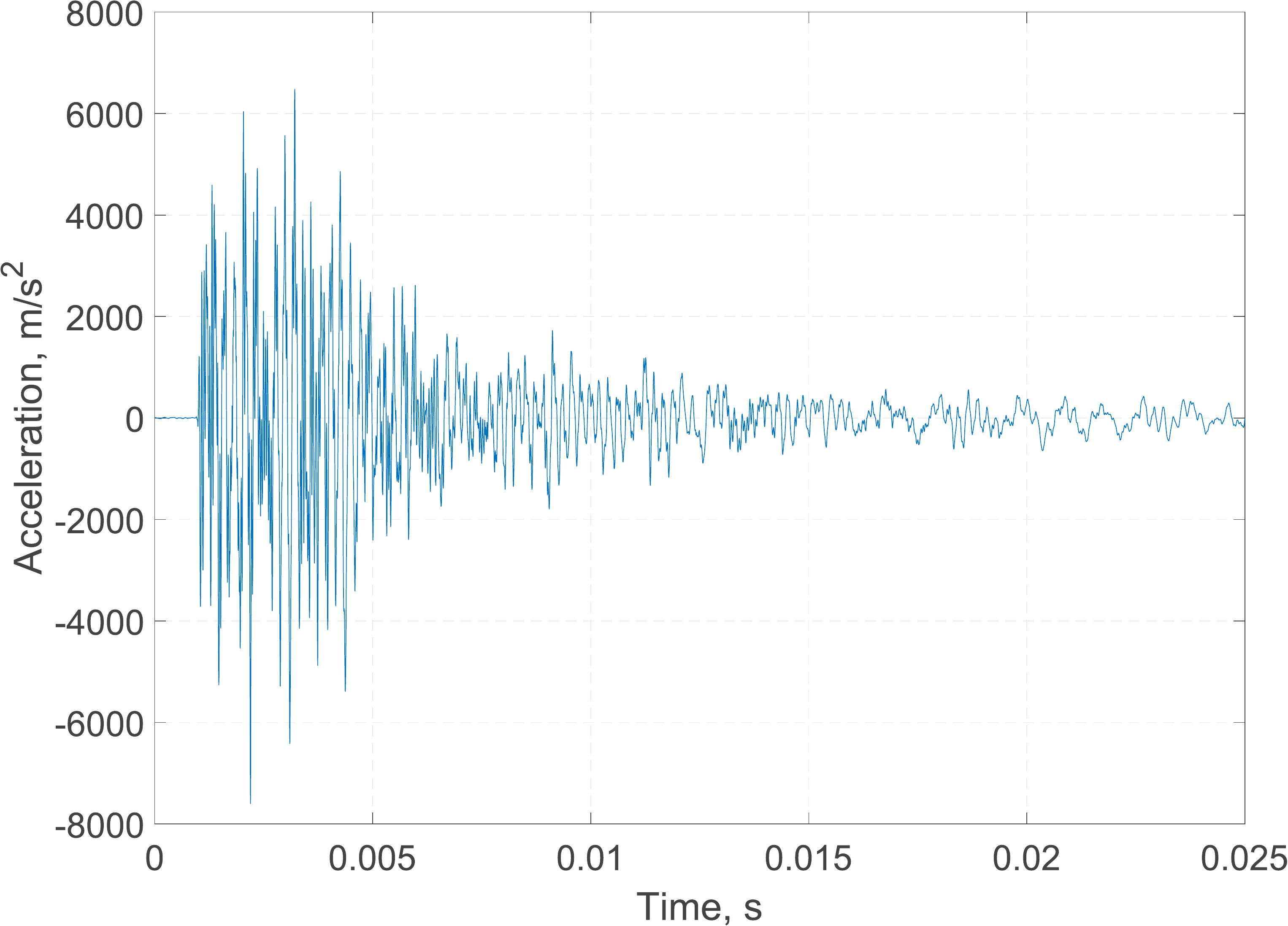}
	\caption{Time history of field shock measurement $\ddot{\u}$}
	\label{ddot_v}
\end{figure}

\begin{figure}
	\centering
	\begin{subfigure}{0.49\textwidth}
		\includegraphics[width=\linewidth]{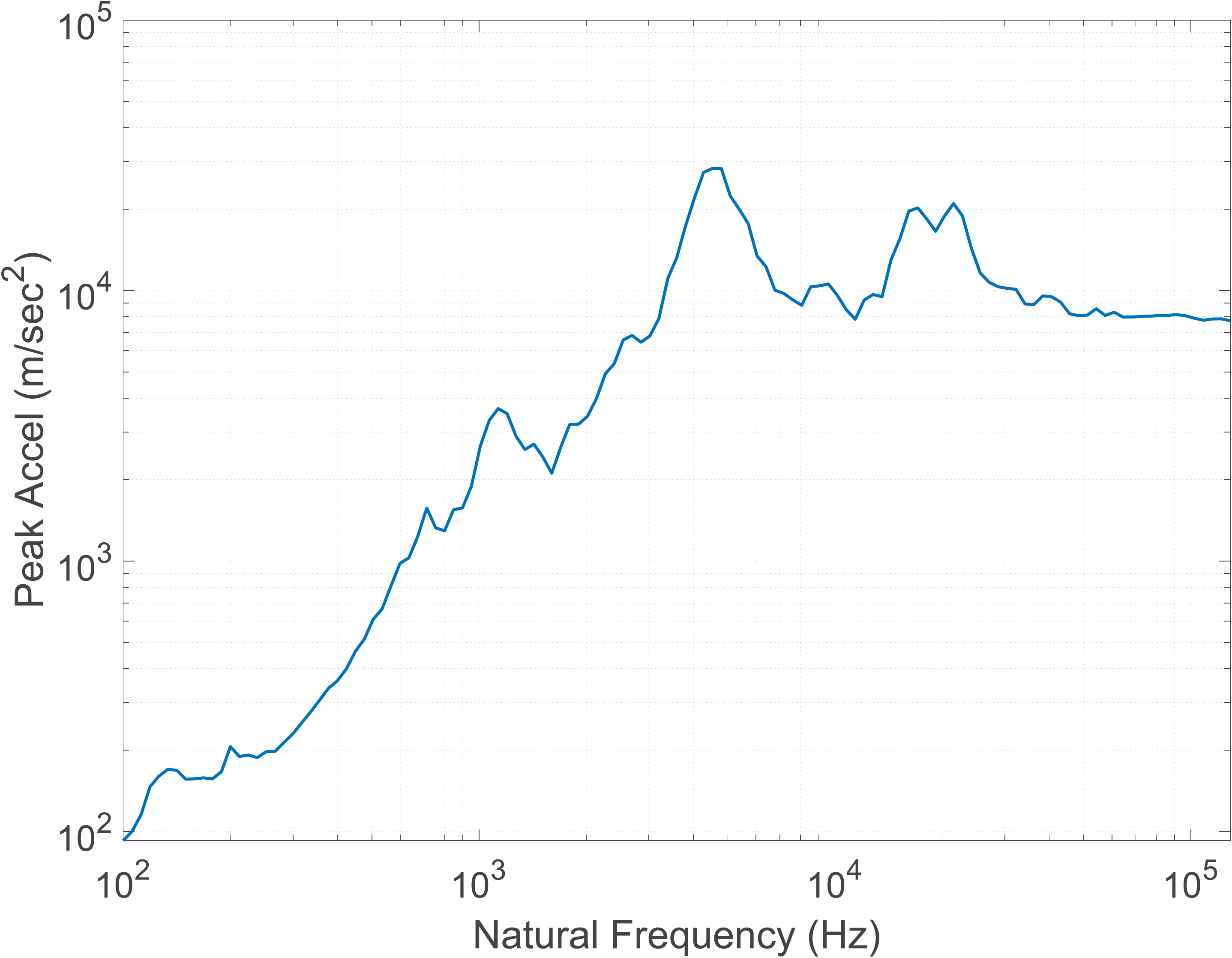}
		\caption{Shock response spectrum}
		\label{measurement_srs}
	\end{subfigure}
	\begin{subfigure}{0.49\textwidth}
		\includegraphics[width=\linewidth]{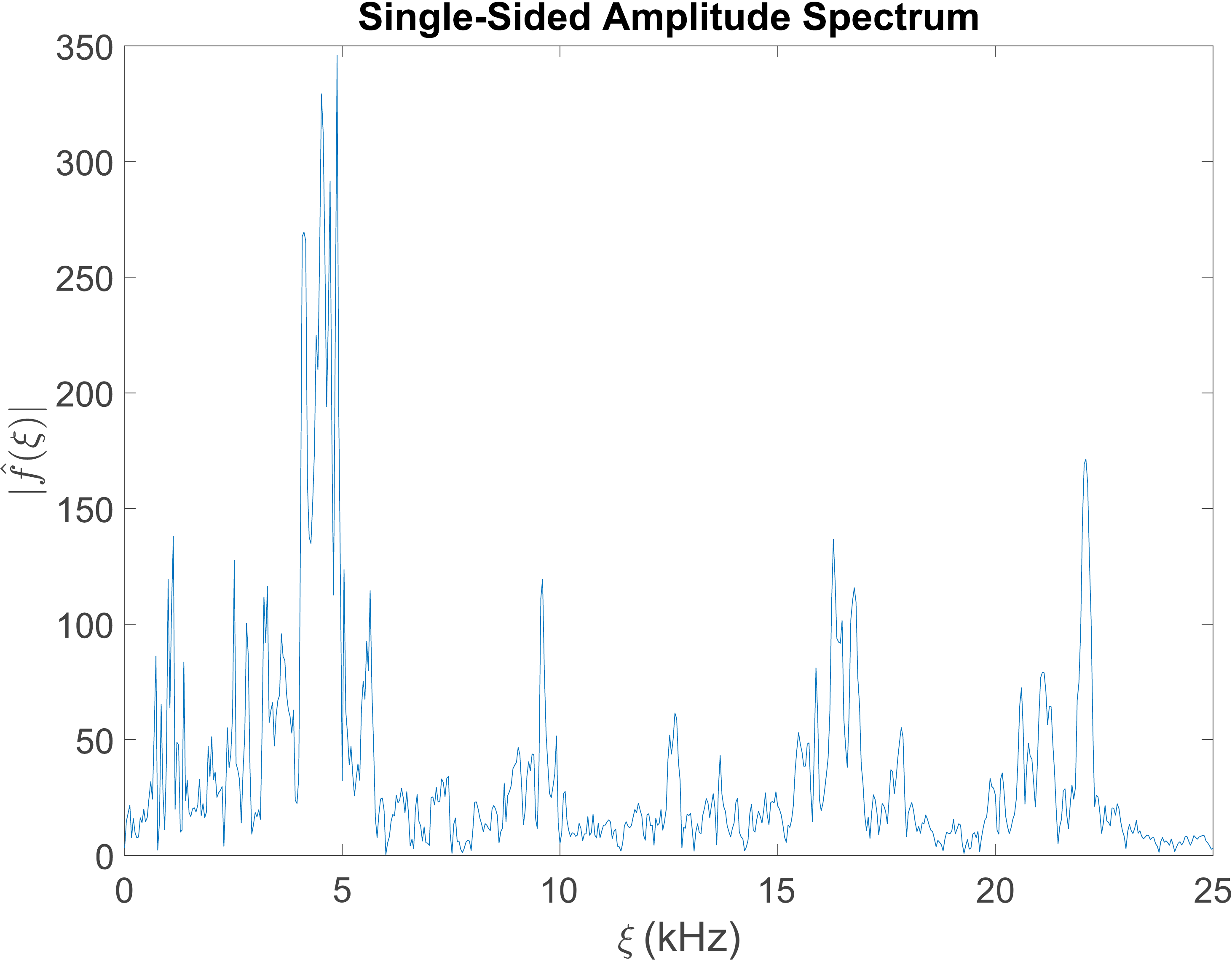}
		\caption{Frequency spectrum}
		\label{measurement_fft}
	\end{subfigure}
	\caption{SRS and frequency spectrum of the field shock measurement $\ddot{\u}$ in Fig.\ref{ddot_v}}
	\label{measurement_srs_fft}
\end{figure}

In this section, two time histories of shocks are synthesised to meet given SRS examples within $\pm$3dB tolerance.
The first example is to synthesize a shock $\ddot{\u}'_{1}$ meeting a typical testing specification in Table \ref{SRS_spec}, which is frequently referred in ESA's mechanical shock handbook\cite[p.~183]{ECSS2015}.
Such specification is often composed of 6 parameters as shown in Table \ref{SRS_spec}, which define an initial slope, a cut-off frequency and a constant plateau in a logarithm SRS graph.
The second example is to synthesize a shock $\ddot{\u}'_{2}$ meeting a more complex SRS from a launcher-induced shock\cite[p.~48]{ECSS2015}.
Syntheses of time histories meeting these complex SRS curves are to demonstrate the claim that the NZDF algorithm can well match any given shock specifications.
The field shock measurement $\ddot{\u}$ shown in Fig.\ref{ddot_v} is generated by mechanical impact in a laboratory environment, whose experimental set-up can be found in Ref.\cite{yan2019general} for detailed information.
The SRS and frequency spectrum of the field shock $\ddot{\u}$ are given in Fig.\ref{measurement_srs_fft}.
The physical similarity and wide frequency range make the field measurement $\ddot{\u}$ suitable for generating general mechanical shocks.

\begin{figure}
	\centering
	\includegraphics[width=0.7\linewidth]{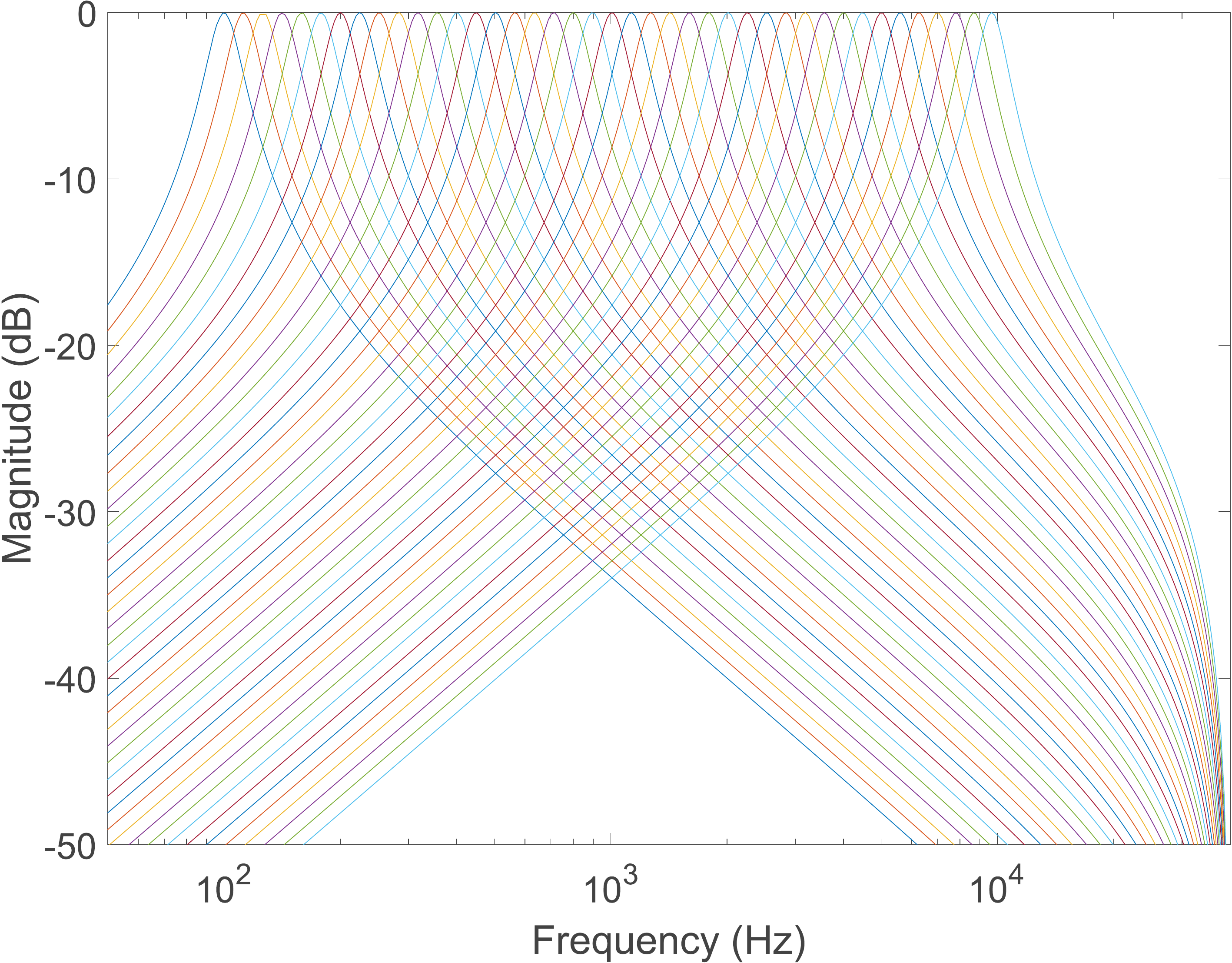}
	\caption{The transfer functions of the NZDF bank in Eq.(\ref{tf_final}) with 1/6 octave centre frequency space}
	\label{gtfb_fpeaks}
\end{figure}

\begin{figure}
	\centering
	\includegraphics[width=\linewidth]{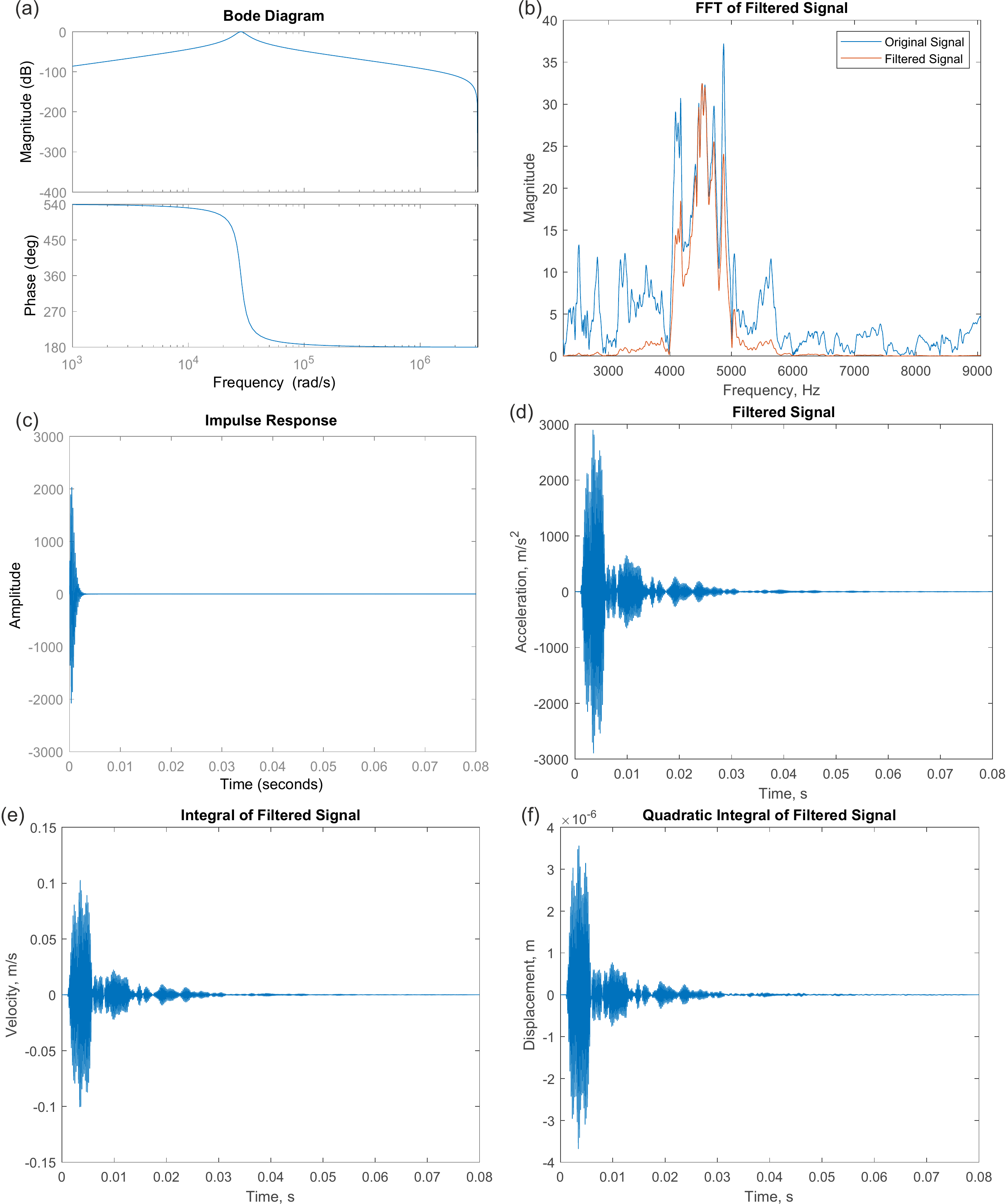}
	\caption{Field shock measurement $\ddot{\u}$ filtered at 4525.5 Hz; (a) Bode diagram of NZDF; (b) comparison of filtered and original signal in frequency spectrum; (c) impulse response of NZDF; (d) acceleration of filtered signal; (e) velocity of filtered signal; (f) displacement of filtered signal.}
	\label{filtered_result}
\end{figure}

A series of NZDFs are generated from Eq.(\ref{tf_final}).
Fig.\ref{gtfb_fpeaks} shows the transfer functions of the filter bank.
The spacing between centre frequencies $\omega_{i}$ is 1/6 octave, which spans the whole frequency range of SRS specification.
To demonstrate the performance of a single NZDF (e.g. $\omega_{34}$ at 4525.5 Hz), Fig.\ref{filtered_result} shows the corresponding Bode plot, frequency spectrum, impulse response, filtered signals in acceleration, velocity and displacement.
The filtered signals in acceleration, velocity and displacement all approach zero when the shock event is finished.
They have almost the same waveform but with different amplitudes and satisfy the following relationship, i.e.,
\begin{equation}\label{4cp}
\ddot{\u}(\omega_i)=\omega_i \times \dot{\u}(\omega_i)=\omega_i^2 \times \u(\omega_i)
\end{equation}
which is the same as that used to construct 4-coordinate graph for earthquake and shock response analysis\cite{gaberson2012shock, newmark1960effect, li2018damage}.

The filtered signals $\ddot{\u}(\omega_{i})$ are normalized by their maximum amplitude and then assembled into the basis matrix $\A$, which are passed to the PSO algorithm to find out a coefficient vector $\x$.
This optimization process usually takes only dozens of seconds if sample points of $\ddot{\u}$ are less than 10,000, e.g., about 4000 sample points are synthesised to meet the SRS specification within 10 seconds.
The computing time may increase to several minutes if more sample points are synthesised.

\begin{figure}
	\centering
	\includegraphics[width=\linewidth]{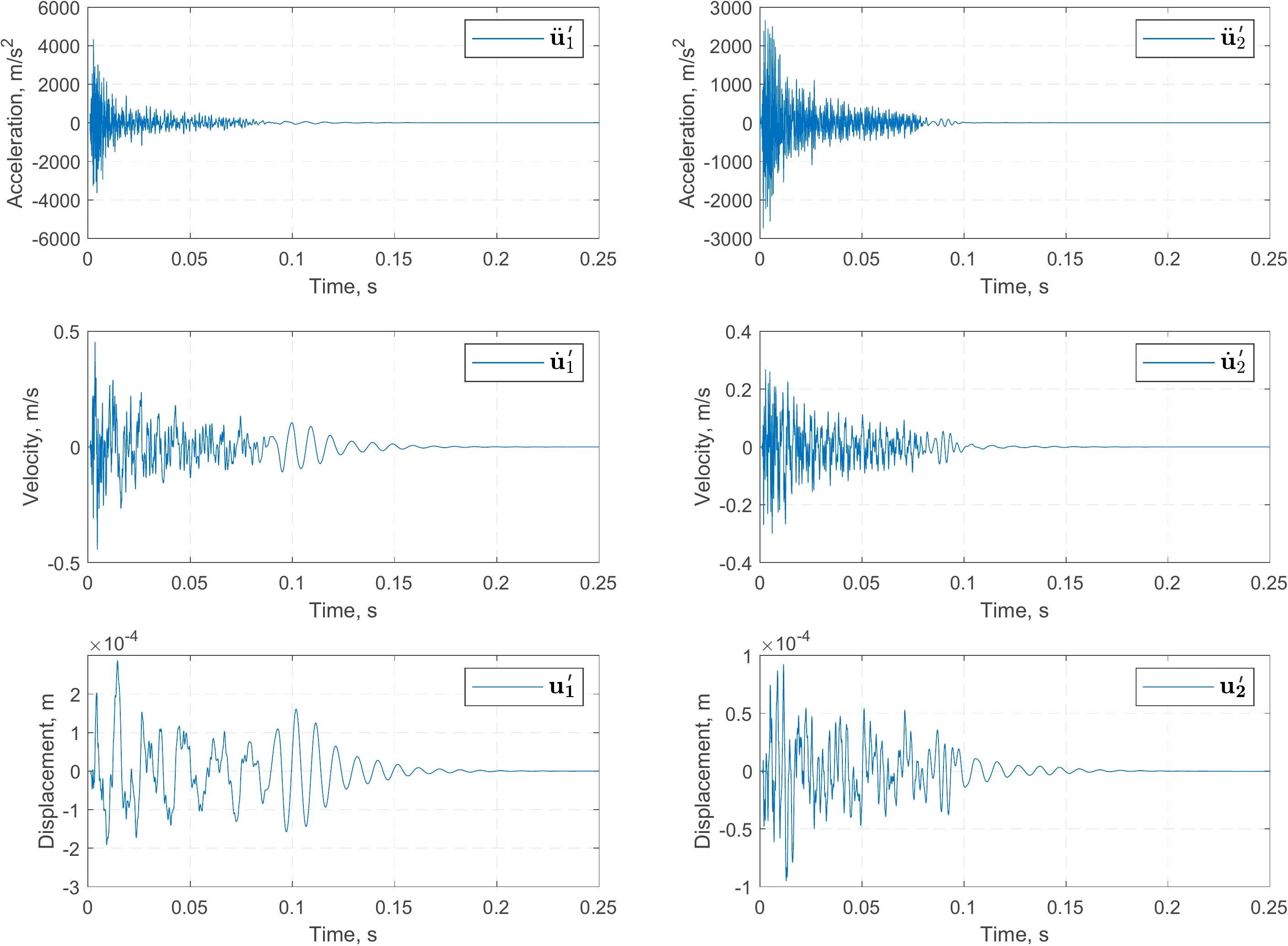}
	\caption{Accelerations $\ddot{\u}'$, velocities $\dot{\u}'$, and displacements $\u'$ of synthesized shocks for both typical testing specification and SRS of launcher-induced shock}
	\label{results_time_history}
\end{figure}

\begin{figure}
	\centering
	\begin{subfigure}{0.49\linewidth}
		\includegraphics[width=\linewidth]{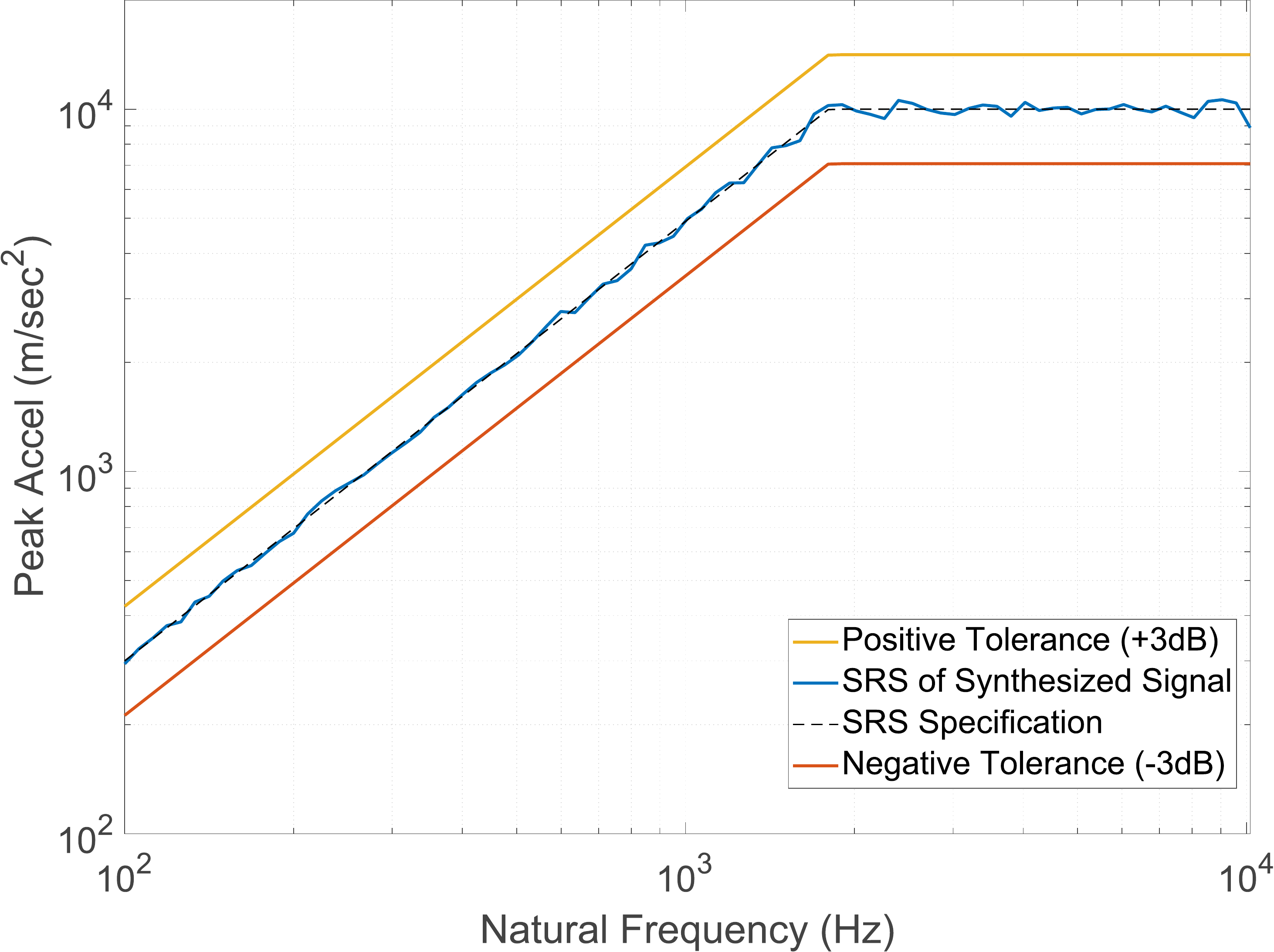}
		\caption{SRS of the typical testing specification}
	\end{subfigure}
	\begin{subfigure}{0.49\linewidth}
		\includegraphics[width=\linewidth]{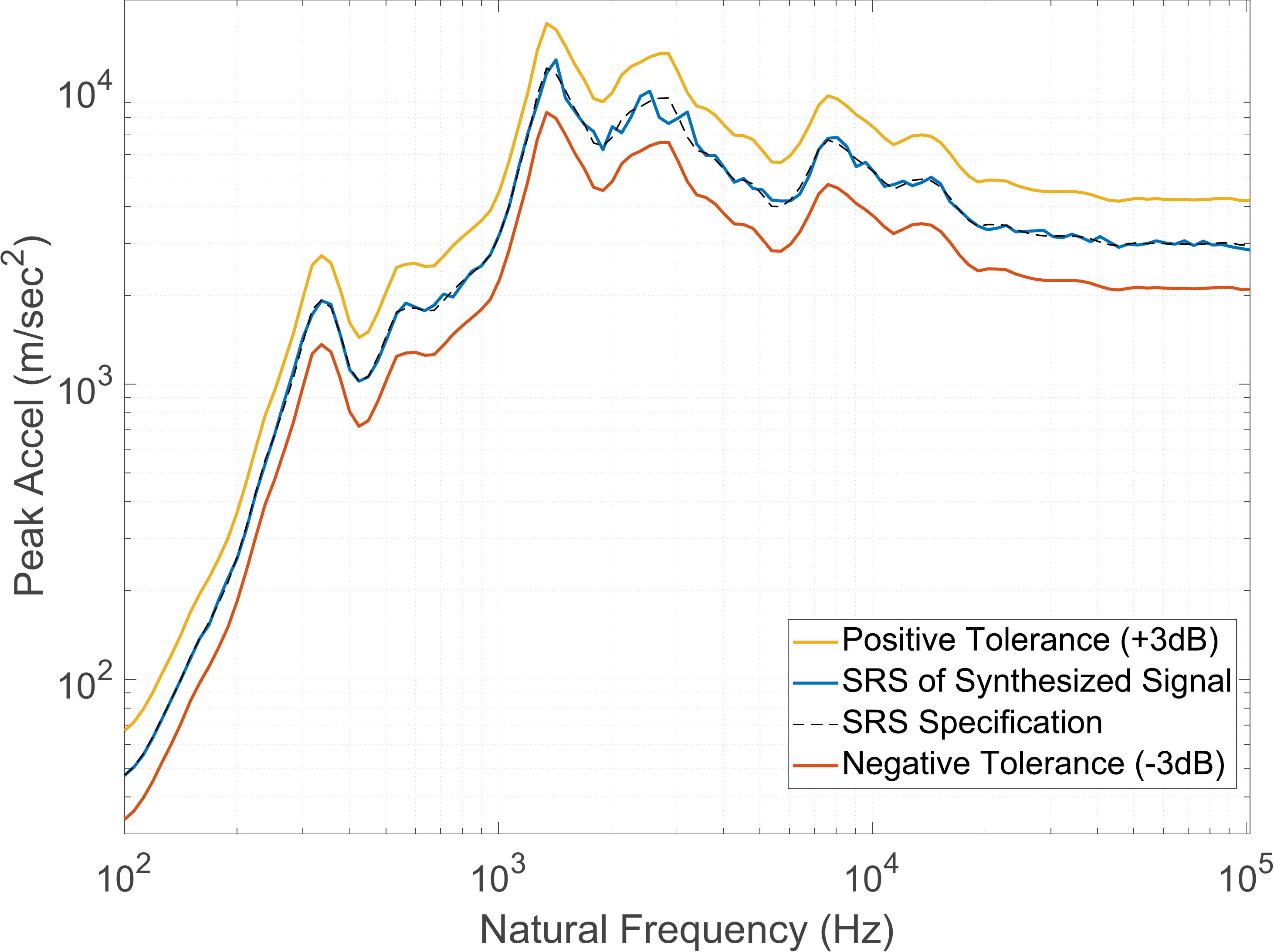}
		\caption{SRS of the launcher-induced shock}
	\end{subfigure}
	\caption{SRS of synthesised shock and specification with $\pm$3dB tolerance}
	\label{result_srs}
\end{figure}

Fig.\ref{results_time_history} shows the accelerations $\ddot{\u}'$, velocities $\dot{\u}'$ and displacements $\u'$ of the synthesised shocks for both the typical testing specification and the SRS of launcher-induced shock.
The synthesised shocks satisfy the net zero displacement change requirement strictly and resemble the field shock measurement $\ddot{\u}$ in terms of its temporal features.
Fig.\ref{result_srs} compares the SRS curves of synthesized shocks and the corresponding specification with $\pm$3dB tolerance.
The SRS curves of synthesized shocks match their corresponding specifications very well, with the error within $\pm$1dB tolerance.

\section{Conclusions}

This paper proposed a class of filters that have a concise expression and can satisfy the net zero displacement change condition.
A set of parameters for the filter bank design are also obtained according to the temporal and frequency characters of general mechanical shock signals.
A shock with net zero displacement change can be synthesised by the linear combination of the filtered field measurement at various centre frequencies.
The SRS of synthesised shocks can well match any given shock specification.

\appendix

\bibliographystyle{elsarticle-num} 
\bibliography{paper_synthesis}

\begin{thebibliography}{10}
\expandafter\ifx\csname url\endcsname\relax
  \def\url#1{\texttt{#1}}\fi
\expandafter\ifx\csname urlprefix\endcsname\relax\def\urlprefix{URL }\fi
\expandafter\ifx\csname href\endcsname\relax
  \def\href#1#2{#2} \def\path#1{#1}\fi

\bibitem{ECSS2015}
ECSS, {Mechanical Shock Design and Verification Handbook}, ESA, Noordwijk,
  Netherlands, 2015.

\bibitem{810g}
{Department of Defence Test Method Standard}, MIL-STD-810G Environmental
  Engineering Considerations and Laboratory Tests, Washington D.C, USA, 2008.

\bibitem{901e}
{Department of Defence Test Method Standard}, MIL-S-901D Shock Tests, H.I.
  (High-Impact) Shipboard Machinery, Equipment, and Systems, Requirements,
  Washington D.C, USA, 2008.

\bibitem{lalanne2013mechanical}
C.~Lalanne, Mechanical Vibration and Shock Analysis, Mechanical Shock, Vol.~2,
  John Wiley \& Sons, Chichester, UK, 2013.

\bibitem{smallwood1975time}
D.~Smallwood, Time history synthesis for shock testing on shakers, Seminar on
  Understanding Digital Control and Analysis in Vibration Test Systems (1975)
  23--42.

\bibitem{kern1984transient}
D.~Kern, C.~Hayes, Transient vibration test criteria for spacecraft hardware,
  The Shock and Vibration Bulletin 54 (1984) 99--109.

\bibitem{fisher1977digital}
D.~Fisher, M.~Posehn, Digital control system for a multiple-actuator shaker,
  The Shock and Vibration Bulletin 3 (1977) 79--96.

\bibitem{yang1972development}
R.~Yang, H.~Saffell, Development of a waveform synthesis technique, The Shock
  and Vibration Bulletin 2 (1972) 45--53.

\bibitem{brake2011inverse}
M.~R. Brake, An inverse shock response spectrum, Mechanical Systems and Signal
  Processing 25~(7) (2011) 2654--2672.

\bibitem{hwang2016stochastic}
J.~H.-J. Hwang, A.~Duran, Stochastic shock response spectrum decomposition
  method based on probabilistic definitions of temporal peak acceleration,
  spectral energy, and phase lag distributions of mechanical impact pyrotechnic
  shock test data, Mechanical Systems and Signal Processing 76 (2016) 424--440.

\bibitem{monti2017dynamic}
R.~Monti, P.~Gasbarri, Dynamic load synthesis for shock numerical simulation in
  space structure design, Acta Astronautica 137 (2017) 222--231.

\bibitem{irvineShock}
T.~Irvine, Shock response spectrum synthesis via wavelets,
  https://vibrationdata.wordpress.com (Accessed 15/03/2019).

\bibitem{Ferebee2008a}
R.~C. Ferebee, J.~Clayton, D.~Alldredge, T.~Irvine, {An Alternative Method of
  Specifying Shock Test Criteria}, Tech. rep., NASA, Huntsville, USA (2008).

\bibitem{yan2019low}
Y.~Yan, Q.~M. Li, Low-pass-filter-based shock response spectrum and the
  evaluation method of transmissibility between equipment and sensitive
  components interfaces, Mechanical Systems and Signal Processing 117 (2019)
  97--115.

\bibitem{katsiamis2007practical}
A.~G. Katsiamis, E.~M. Drakakis, R.~F. Lyon, Practical gammatone-like filters
  for auditory processing, EURASIP Journal on Audio, Speech, and Music
  Processing 2007~(1) (2007) 063685.

\bibitem{lyon2017human}
R.~F. Lyon, Human and Machine Hearing, Cambridge University Press, Cambridge,
  UK, 2017.

\bibitem{yan2019general}
Y.~Yan, Q.~M. Li, {A general shock waveform and characterisation method},
  Mechanical Systems and Signal Processing 136 (2020) 106508.

\bibitem{flanagan1960models}
J.~L. Flanagan, Models for approximating basilar membrane displacement, Bell
  System Technical Journal 39~(5) (1960) 1163--1191.

\bibitem{johannesma1972pre}
P.~I.~M. Johannesma, {The pre-response stimulus ensemble of neurons in the
  cochlear nucleus}, in: Symposium on Hearing Theory, Eindhoven, Netherlands,
  1972.

\bibitem{aertsen1980spectro}
A.~M. H.~J. Aertsen, P.~I.~M. Johannesma, Spectro-temporal receptive fields of
  auditory neurons in the grassfrog, Biological Cybernetics 38~(4) (1980)
  223--234.

\bibitem{lyon1997all}
R.~F. Lyon, All-pole models of auditory filtering, Diversity in auditory
  mechanics (1997) 205--211.

\bibitem{slaney1993efficient}
M.~Slaney, {An efficient implementation of the Patterson-Holdsworth auditory
  filter bank}, Tech. Rep.~35, Apple Computer Technical Report, Cupertino, USA
  (1993).

\bibitem{gaberson2012shock}
H.~A. Gaberson, Shock severity estimation, Sound \& Vibration 46~(1) (2012)
  12--20.

\bibitem{newmark1960effect}
N.~M. Newmark, {Effect of inelastic behavior of the response of simple systems
  to earthquake motions}, in: Proceedings of the 2nd World Conference on
  Earthquake Engineering, Tokyo, Japan, 1960.

\bibitem{li2018damage}
B.~Li, Q.~M. Li, Damage boundary of structural components under shock
  environment, International Journal of Impact Engineering 118 (2018) 67--77.

\end{thebibliography}

\section{Proof of Eq.(\ref{m_bound})}\label{proof_appendix}

Substituting Eq.(\ref{displacement_laplace}) to Eq.(\ref{initial_initial_theorem}), we have
\begin{equation}\label{simplified_final_value}
\lim\limits_{s \rightarrow 0} \frac{ K s^{M-1} \ddot{U}(s)}{(s^2 +(\omega/Q)s + \omega^2)^N}
=\frac{K}{\omega^{2N}} \cdot \lim\limits_{s \rightarrow 0} s^{M-1}\ddot{U}(s)
=0
\end{equation}
According to Laplace transform,
\begin{equation}
\ddot{U}(0)=\int_{0}^{\infty} \ddot{u}(t) \ dt,
\end{equation}
which is usually a bounded finite value.
Therefore, Eq.(\ref{simplified_final_value}) is equivalent to $\lim\limits_{s \rightarrow 0} s^{M-1}=0$,
\begin{flalign}
\text{i.e.,}&&M \geq 2&&
\end{flalign}
Substituting Eq.(\ref{tf_NZDF}) into Eq.(\ref{initial_initial_theorem}), we have
\begin{equation}
\lim\limits_{s \rightarrow \infty} \frac{ K s^{M+1} }{(s^2 +(\omega/Q)s + \omega^2)^N}.
\end{equation}
By L'Hôpital's rule, this limit is finite only if
\begin{flalign}
&&M+1 \leq 2N&&\\
\text{or}&&M \leq 2N-1&.&
\end{flalign}

\end{document}